\documentclass[10pt]{article}

\usepackage{amsmath}
\usepackage{amssymb}
\usepackage{amsthm}
\usepackage{colonequals}
\usepackage{enumerate}

\newtheorem{definition}{Definition}

\newtheorem{remark}{Remark}

\DeclareMathOperator{\argmax}{{\mathrm{argmax}}}
\newcommand{\calN}{\mathcal{N}}

\begin{document}

\title{Subjective Collaborative Filtering}
\author{
Fabrizio Caruso \\ Neodata Group, Catania \\ \texttt{fabrizio.caruso@neodatagroup.com} \and 
Giovanni Giuffrida \\ Dept. of Social Sciences, University of Catania \\ \texttt{ggiuffrida@dmi.unict.it} \and 
Calogero Zarba \\ Neodata Group, Catania \\ \texttt{calogero.zarba@neodatagroup.com}}

\maketitle

\begin{abstract}

We present an item-based approach for collaborative filtering.
We determine a list of recommended items 
for a user by considering their previous
purchases. Additionally other features of the users
could be considered such as page views, search queries, etc\dots

In particular we address the problem of 
efficiently comparing items.
Our algorithm can efficiently
approximate an estimate of the similarity between
two items. 
As measure of similarity
we use an approximation of the Jaccard similarity 
that can be computed by
constant time operations and one bitwise \texttt{OR}.

Moreover we improve the accuracy of the similarity by introducing
the concept of user preference for a given product, which
both takes into account multiple purchases and purchases
of related items.
The product of the user preference and the Jaccard measure
(or its approximation) 
is used as a score for deciding whether a given product 
has to be recommended.

\medskip


\end{abstract}

\section{Introduction}

We have a web site where a list of items are offered for purchase.
We can also assume that a user can perform other actions such as  
clicking on advertisement banners, reading a review of a product, 
querying the system with a keyword, etc\dots
 
Our goal is to produce the best possible suggestion to a user
depending primarily on his previously bought items (if any) and
secondarily on his behavior.

For a given user $u$, we store all events of the user $u$ 
in the cookie maintained by
the browser.  The cookie contains also the timestamps of the
stored events.
The user's cookie is used by the webserver each time the user requests
a webpage, in order to select an appropriate suggestion.  

Two approaches based on similarity are commonly followed 
in collaborative filtering:
one based on similar items (known as item-based or model-based)
and one based on similar users (known as user-based or memory-based).
Two products are similar if they have been bought
by common users.
Two users are similar if they have bought common
items.
Different measures of similarities have been used
in the literature
such as Jaccard similarity, cosine similarity, 
Pearson correlation index, etc\dots\

In this article we present an item-based approach that
addresses the issue of scalability
by efficiently approximating the similarities among
items (possibly in real time).
For a justification of an item-based approach we
refer to~\cite{Amazon} and~\cite{Sarwar:2001}.
The main ideas for the efficiency of the algorithm 
is the application of the linear counting algorithm~\cite{WhangZandenTaylor1990}
for the approximation of the Jaccard measure of similarity
among items.

In the simplest version of our approach we are only considering 
the implicit binary rating, i.e.,
whether an item has been bought at least once by a user and we are 
disregarding the quantity. 
An extension of our approach to non-Boolean purchases
in described in Section~\ref{se:non-Boolean}.

Efficiently estimating the value f the Jaccard similarity can be a
daunting task when the number of (usually non-unique) users in the
webserver's log can reach huge figures (millions).  We propose an
approximate estimation based on the linear counting
algorithm~\cite{WhangZandenTaylor1990}, i.e., approximately counting
the number of unique users that have bought a specific item.

We improve the accuracy of the similarity by introducing
the concept of user preference for a given product, which
both takes into account multiple purchases and purchases
of related items.
The product of the user preference and the Jaccard measure
(or its approximation) 
is used as a score for deciding whether a given product 
has to be recommended.

The paper is organized as follows:
In Section~\ref{se:pre} we introduce some notation and definitions.
In Section~\ref{se:lc} we present the approach based on 
approximately counting unique users.
In Section~\ref{se:non-Boolean} we describe how to extend the
algorithm to the non-Boolean case.
In Section~\ref{sse:behavior} we consider a generalization of the
approach that takes into account
the user's behavior (page views, search queries, clicks on banners, etc\dots).
In Section~\ref{sse:ad} we consider a possible adaptation of our
approach to ad-serving.

\section{Preliminaries}\label{se:pre}
We will be using two data-structures to store information on
user's actions at real time: 
a user's {\em profile}, which depends on the user;
a {\em purchase matrix}, which is global.

In this article we will denote with $P$ the set of all products
and by $U$ the set of all users.

The next definition describes the data that has to
be stored by each user's cookie.

\begin{definition}[User's profile]
For every user $u$ we define their {\em profile} $P_u$
which is simply the subset of $P_u \subseteq P$ containing
all their purchased items.
\end{definition}

In the following definition we describe the
global data-structure necessary in our approach.

\begin{definition}
We denote by $M=(m_{p,u})$ the ``purchase matrix'',
a $\text{product} \times \text{user}$ matrix
where $m_{p,u}$ is the number of purchases of
product $p$ by user $u$.
\end{definition}

\section{Recommending by counting unique users}
\label{se:lc}
Let us consider an item-based approach in which we can
disregard the number of purchases of a given
item.
We assume that our recommending system produces
a given number of suggestions to any user who has bought at
least one item.

Then we can estimate the similarity of two items
$p$ and $c$
through the Jaccard measure:
\begin{equation}\label{eq:Jaccard}
J(p,c) = \frac{|U_p \cap U_c|}{|U_p \cup U_c|},
\end{equation}
\noindent where $U_p$ and $U_c$ are the set of users
that have bought (at least once) $p$ ($p$-th column of 
the matrix $M$) and $c$ ($c$-th column of $M$), respectively.
\begin{remark}
The Jaccard similarity between a previously bought item $p$
and a candidate item $c$ describes an ``objective similarity''
between $p$ and $c$. In particular $J(p,c)$ does not depend
on the specific target user we are considering since it is
applied without any change to any user that has bought $p$ and not $c$. 
\end{remark}

We denote by $\calN(p)$ a function that maps a product $p \in P$
into its set of ``neighbors'', i.e. most similar products.
Possible choices for $\calN(p)$ could be the set of $k$-nearest neighbors
or a set of products such that their Jaccard similarity is above
a given threshold.
We define $\calN^{+}(p):= \calN(p) \cup \{ p \}$.

We denote by $C_u$ the set of candidate products for user $u$.
Some possible choices for $C_u$ are $\calN(p)$ and $P \setminus P_u$.

\subsection{Off-line computing the similarities}

As discussed in \cite{Amazon}, \cite{Sarwar:2001}, similarities
among items tend to change slowly and much more slowly than 
similarities among users.
This justifies an off-line computation of the similarities $J(p,c)$
(e.g., daily, hourly) among all possible couples $(p,c)$ of products.
The complexity of the off-line computation of the similarities is given by
$\mathcal{O}(|P|^2 \cdot J)$, where $J$ is the complexity of computing
a single Jaccard similarity.
In practice, most users only buy few
items and the real complexity is similar to $\mathcal{O}(|P| \cdot J)$

\subsection{The ``objective'' algorithm}
\begin{enumerate}[I.]
\item We precompute $J(p,q)$ for any possible couple $(p,q)$
of products.
\item For any given user $u$ we
recommend $c$ taking by
\begin{equation}\label{eq:best-fit}
\underset{c \in C_u}{\argmax}  \max_{p \in P_u} J(p,c).
\end{equation}
\end{enumerate}

Thus, if $J(p,q)$ is computed off-line, the time complexity of 
the above formula is given by $\mathcal{O}(|C_u| \cdot |P_u|)$.

\subsection{Approximating the similarity}
We consider the problem to efficiently approximate the ratio in
equation~(\ref{eq:Jaccard}).
We denote with $U_p$ the set of unique users that
have bought product $p$.
This estimate can be computed using the linear counting
algorithm~\cite{WhangZandenTaylor1990}.  
Thus, for each product $p$, a
bitvector $v_p$ of size $m$ (much smaller than $|U|$) is kept, where $m$ is a sufficiently 
large constant.  For each purchase, the user is hashed into a bucket
of the bitvector $v_p$, which is then set to $1$.  The estimate of the
number of unique users that have bought product $p$ is then given
by the formula $-m \log(\hat{v}_p)$, where $\hat{v}_p$ is the ratio of
$0$ bits in the bitvector $v_p$.
Through this algorithm we can approximate $|U_p|$ and $|U_c|$ in 
equation~(\ref{eq:Jaccard}).
We can use this procedure to estimate the ratio in equation~(\ref{eq:Jaccard}),
in that we estimate $|U_p \cup U_c|$ and $|U_p \cap U_c|$.
We approximate $|U_p \cup U_c|$ by taking
$\hat{w}= v_p \text{ \sc or } v_c$ (bitwise $\text{ \sc or}$ of $v_{p}$ 
and $v_{c}$). Then $|U_p \cup U_c|$ is estimated by $-m \log(\hat{w})$.  
Finally, since $|U_p \cup U_c| = |U_p| + |U_c| - |U_p \cap U_c|$ 
we can estimate $|U_p \cap U_c|$ by 
$-m \log(\hat{v}_{p}) - m \log(\hat{v}_{c}) + m \log(\hat{w})$.

Therefore the complexity of the off-line computation of the approximated similarities is given by
$\mathcal{O}(|P|^2 \cdot m)$, and in practice $\mathcal{O}(|P| \cdot m)$.
We remark the $m$ is much smaller than the set $|U|$ of all users
(the suggested value in \cite{WhangZandenTaylor1990} for $m$ is about $|U|/10$).
This is an improvement over other item-based approaches
such as the one proposed in \cite{Amazon}.

\section{Non-Boolean purchases}\label{se:non-Boolean}
This approach, although inherently Boolean, can be generalized in a way that takes into
account both multiple purchases of an item and purchases of similar items.
In this section we describe three ways to achieve this: 
by a score that takes the quantity of purchases into account,
by considering clusters of similar bought items,
by taking similar items into account without explicitly computing clusters.

\subsection{Counting the purchases of an item}\label{sse:user}
The Jaccard measure $J(p,c)$ between a previously
bought item $p$ and candidate a item $c \in \calN(p)$ can be corrected
by a factor that depends on the quantity of item~$p$.  
A possible factor
could be the ratio among the quantity and the average quantity over all purchases
of item~$p$. A probably better factor is proposed in Remark~\ref{re:ranking}.

\subsection{Clustering similar previously bought items}\label{sse:clustering}
We can cluster very similar items:
if two items are very similar (e.g., they differ by a few users)
their sum as vectors could be considered instead of them.

\subsection{Subjective similarity}
We can improve the approach in Section~\ref{sse:user}
and in Section~\ref{sse:clustering}
by considering a more general concept of user preference
$\gamma_u(p)$ for an item $p$ by user $u$ in that we
also consider
the {\em user preference} for related items 
(similar items, items similar to similar items, etc\dots).

\begin{definition}
Given a user $u$ and a candidate item $c$ we define
the {\em subjective similarity} $S_u(p,c)$ between a previously purchased
item $p$ and a candidate item $c$ for $u$ as follows:
\begin{equation}
S_u(p,c) = \gamma_u(p) \cdot J(p,c),
\end{equation}
\noindent where $\gamma_u(p)$ is a measure of the preference for product $p$ by user $u$.
\end{definition}

By doing so we are introducing a ``subjective'' element when comparing $p$ and $c$,
in that we take the user preference for $p$ into account.

\subsection{User preference}
As $\gamma_u(p)$ we could simply take $\gamma_u^{(0)}(p):=m_{p,u}$ (this is exactly the approach in 
Section~\ref{sse:user}), or take into account the preference
for similar items $\gamma_u(p)^{(1)}$: 
\begin{equation}\label{eq:gamma}
\begin{split}
\gamma_u^{(1)}(p) &:= \sum_{q \in \calN^{+}(p)}( m_{q,u} \cdot J(p,q)) = \\
&= m_{p,u} + \sum_{q \in \calN(p)}( m_{q,u} \cdot J(p,q)),
\end{split}
\end{equation}
\noindent in which $m_{q,u}$ is the $u$-th component in $q$, i.e. the quantity of item $q$
bought by user $u$ (normalized with respect to the quantity bought by 
the average user). 

\smallskip

This can be generalized in a recursive fashion (similar to the popular
PageRank algorithm \cite{PageRank}) such that $t$ indirect
similarities of products are considered:
\begin{equation}
\gamma_u^{(t)}(p) := 
\sum_{q \in \calN^{+}(p)}  \gamma_u^{(t-1)}(q) J(p,q),
\end{equation}
\noindent where $\gamma_u^{(0)}(q) = m_{q,u}$.

\smallskip

We can unroll this recursive formula for $\gamma_u^{(t)}(p)$ into the following expanded form:
\begin{equation*}
\sum_{p^{(1)}\in \calN^{+}(p)}
\cdots \sum_{p^{(t)} \in \calN^{+}(p^{(t-1)})}
m_{p^{(t)},u} \prod_{i=1}^{t-1} J(p^{(i-1)},p^{(i)}).
\end{equation*}

\begin{remark}\label{re:ranking}
We can further improve this approach by considering a
slower than linear increasing function $\rho$ that better models
how a quantity should implicitly correspond to a ranking,
than simply taking the normalized quantity $\gamma^{(0)}_u(q)=m_{q,u}$ of a purchased
item (which linearly increases with the number of purchases).
For instance, for a given item $q$,
we can take $\gamma^{(0)}_u(q)$ as follows
\begin{equation}
\rho_M(q) = M \sum_{i=1}^{m_{q,u}} 2^{-i},
\end{equation}
\noindent where $M$ is a desired maximum value for the ranking.
\end{remark}

\subsection{The ``subjective'' algorithm}
\begin{enumerate}[I.]
\item We precompute $J(p,q)$ for any possible couple $(p,q)$
of products.
\item For any given user $u$ we
recommend $c$ by the formula:
\begin{equation}\label{eq:best-subj-fit}
\underset{c \in C_u}{\argmax}  \max_{p \in P_u} S_u(p,c).
\end{equation}
\end{enumerate}

Therefore the complexity for the computation of (\ref{eq:best-subj-fit})
is given by $\mathcal{O}(|C_u|\cdot|P_u|\cdot G)$, where
$G$ is the complexity of the computation of $\gamma^{(t)}_u(p)$,
which depends on $t$.
For $t=0$, the complexity of (\ref{eq:best-subj-fit}) is the
same as the one of (\ref{eq:best-fit}).
For $t>1$, the complexity will depend on the size of $\calN(p)$.
For practical purposes we would not suggest $t>2$ because
the improvement in accuracy is minimal and the complexity worsens.

\section{Future work}
We plan to generalize and modify our approach in two directions:
\begin{enumerate}
\item including the general user's behavior into our approach; 
\item adapting the approach to ad-serving.
\end{enumerate}

Both generalizations will require experiments 
on large sets of real data.

\subsection{Behavioral targeting}\label{sse:behavior}
We can take into account 
events different from a purchase such as page views,
clicks on banners, search queries, etc..., which,
together with purchases, we
call ``features''.
This is achieved by having $c$ in (\ref{eq:best-fit})
run over all features of the given user.
Having different types of features poses the problem
of weighing them with respect to their relevance,
e.g.,
a purchase should count more than any other feature,
a search query should count more than a click on a
banner, etc...
In order to properly fine tune this generalized approach
we need to perform experiments on large data-sets.

\subsection{Application to ad-serving}\label{sse:ad}
We can also apply our approach to ad-serving.
This is probably only possible for large data-sets.
We have drawn this conclusion by real experimental data
available at Neodata:
in our data only $1.5\%$ of the users clicks on two different banners.
Therefore a large set of data is necessary in order to be of
any use for a collaborative filtering algorithm.

Our approach could be applied to behavioral
targeting of banner advertisements in which,
instead of suggesting the product with
highest likelihood of being bought,
the banner
advertisement with highest probability
of being clicked or the one
generating the highest average profit
is displayed.

Nevertheless we cannot use these approach for
ad-serving in a straightforward way
because
buying an item and clicking on a banner
are events of different nature:
the former often depends on the user's
desire of buying an item;
the latter is often a random event
strongly dependent on the number of previous
impressions of the banner.

We can overcome this by having first a learning
phase in which all banners are equally shown
to all users, second using
our approach to compute the banner with
generates the highest profit, i.e. with the
highest ecpm.
An alternative solution may be to normalize
our approach by the number of impressions of the
given banner.

\bibliographystyle{plain}

\end{document}